\documentclass[superscriptaddress, reprint, amsmath, amssymb, aps, pra, floatfix]{revtex4-2}

\usepackage{bm}
\usepackage{mathrsfs}
\usepackage{algorithm}
\usepackage{algpseudocode}
\usepackage{tocloft}
\newcommand{\listappendicesname}{Appendix Contents}
\newlistof{appendix}{app}{\listappendicesname}
\usepackage{caption}

\usepackage{ragged2e}
\usepackage{booktabs}
\usepackage{blindtext}
\usepackage[skip=0pt]{caption}
\usepackage{comment}
\usepackage{mathtools}
\usepackage{mathrsfs}
\usepackage{amsfonts}
\usepackage{amssymb}
\usepackage{dsfont}
\usepackage{amsthm}
\usepackage[english]{babel}
\usepackage{graphicx}
\usepackage{amsmath}
\usepackage{lipsum}
\usepackage{wrapfig}
\usepackage{float}
\usepackage{enumitem} 
\usepackage{color}
\usepackage{hyperref}
\usepackage[normalem]{ulem}
\usepackage{url}
\usepackage{subcaption}
\usepackage{ragged2e}

\useunder{\uline}{\ul}{}

\hypersetup{colorlinks = true}

\def\bra#1{\ensuremath{\mathinner{\langle{#1}|}}}
\def\ket#1{\ensuremath{\mathinner{|{#1}\rangle}}}
\newcommand{\norm}[1]{\left\lVert #1 \right\rVert}

\newcommand{\needcite}[1]{\textcolor{red}{[Ref needed]}}

\newcommand{\nocontentsline}[3]{}
\newcommand{\tocless}[2]{\bgroup\let\addcontentsline=\nocontentsline#1{#2}\egroup}

\begin{document}

\title{Higher-order Zeno sequences}
\author{Kasra Rajabzadeh Dizaji}

\affiliation{School of Electrical, Computer, and Energy Engineering, Arizona State University, Tempe, Arizona 85287, USA}
\affiliation{Department of Physics, Arizona State University, Tempe, Arizona 85281, USA}
\author{Leeseok Kim}
\affiliation{Department of Electrical \& Computer Engineering and Center for Quantum Information and Control, University of New Mexico, Albuquerque, NM 87131, USA}
\author{Milad Marvian}
\affiliation{Department of Electrical \& Computer Engineering and Center for Quantum Information and Control, University of New Mexico, Albuquerque, NM 87131, USA}
\author{Christian Arenz}
\affiliation{School of Electrical, Computer, and Energy Engineering, Arizona State University, Tempe, Arizona 85287, USA}
\date{\today} 
\begin{abstract}

The quantum Zeno effect typically refers to freezing the dynamics of a quantum system through frequent observations. In general, quantum Zeno dynamics is obtained with an error of order $\mathcal{O}(1/N)$, where $N$ is the number of projective measurements performed within a fixed evolution time. In this work, we develop higher-order Zeno sequences that achieve faster convergence to Zeno dynamics, yielding an improved error scaling of $\mathcal{O}(1/N^{2k})$, where $k$ describes the order of the Zeno sequence. This is achieved by relating higher-order Zeno sequences to higher-order Trotter formulas that achieve similar convergence behavior. We leverage this relation to develop higher-order Zeno sequences for different manifestations of the quantum Zeno effect, including frequent projective measurements and unitary kicks. 
We go on to discuss achieving quantum Zeno dynamics through periodic control fields of high frequency. We explicitly develop control fields that yield a second-order type improvement in the Zeno error scaling and present shorter Zeno sequences. Finally, we discuss the connection to randomized and Uhrig dynamical decoupling to develop more efficient implementations in the weak coupling regime.

\end{abstract}
\maketitle
\tocless\section{Introduction}
The quantum Zeno effect is an intriguing effect in quantum mechanics. Frequent projective measurements can modify the evolution of a quantum system in a way that it becomes entirely frozen \cite{misra1977zeno, facchi2008quantum} or constrained to the (Zeno) subspace in which the dynamics can be more complex \cite{Facchi_2002,eb3d95ca9b7c45eab60553aab2734edb,Arenz_2016}. The quantum Zeno effect has been experimentally observed in various systems such as trapped ions \cite{BALZER2002235,Toschek_2001} and photonic systems \cite{PhysRevLett.74.4763, MOLHAVE200045}.

Quantum Zeno dynamics can be achieved through frequent projective measurements, the rapid application of unitary kicks, or through the application of a periodic control field of high frequency \cite{Facchi_2003}. Applications of the quantum Zeno effect include the preservation of spin polarization \cite{PhysRevA.65.013404}, dosage reduction in neutron tomography \cite{PhysRevA.66.012110}, decoherence suppression \cite{PhysRevA.71.022302, PhysRevB.71.165314}, Hamiltonian simulation \cite{dizaji2024hamiltoniansimulationzenosubspaces}, Hamiltonian learning \cite{franceschetto2025hamiltonianlearningquantumzeno}, and quantum control in general \cite{Lewalle_2024,PhysRevA.70.062302}.

The quantum Zeno effect is typically achieved when $N$ projective measurements are performed at a rate $\mathcal{O}(1/N)$ within a fixed evolution time $t$. More specifically, the probability of failing to obtain the system in its initial state at a time $t$ when $N$ projective measurements are performed in time intervals $\Delta t=\frac{t}{N}$ vanishes as $\mathcal{O}(1/N)$ \cite{facchi2008quantum}. In a seminal paper by Grover et al. \cite{dhar2006preserving}, it has been shown that transitions to other states can be suppressed at a faster rate when the observed dynamics is additionally interspersed with the unitary reflection operator $R=P-Q$, where $P$ is the projection describing the measurement and $Q$ is its orthogonal complement. Building on these results, it was recently shown by some of the authors \cite{dizaji2024hamiltoniansimulationzenosubspaces} that such a strategy can also be employed to achieve Zeno dynamics faster with a quadratically improved rate $\mathcal{O}(1/N^{2})$. Furthermore, very recently, higher-order Zeno sequences have been proposed based on linear combinations of previous sequences~\cite{möbus2024multiproductzenoeffecthigher}. However, the implementation of these sequences through unitary operations and projective measurements remains challenging.  

In this work, we generalize the results in \cite{dhar2006preserving,dizaji2024hamiltoniansimulationzenosubspaces} to develop higher-order Zeno formulas that yield convergence to Zeno dynamics at a rate $\mathcal{O}(1/N^{2k})$, where $k$ describes the order of the sequence. This is achieved by relating higher-order Zeno sequences that are constructed through the inclusion of $R$ with higher-order Trotter formulas. In turn, this relation allows leveraging the theory of higher-order Trotter sequences \cite{childs2021theory} to derive error bounds and obtain asymptotic Zeno error scaling estimates for the proposed higher-order Zeno sequences. We explicitly construct such sequences for the Zeno effect achieved with projective measurements and unitary kicks. For the latter, we show that the presented approach yields tighter bounds than the bounds previously reported in the literature \cite{hahn2022unification}. We go on to discuss higher-order Zeno sequence implementations through periodic control fields and explicitly construct control fields that achieve second-order convergence. Finally, we discuss higher-order Zeno sequences that require significantly fewer reflection operations by exploiting their connection to Uhrig dynamical decoupling in the weak-coupling limit \cite{uhrig2007keeping,yang2008universality}, as well as randomized techniques that further improve convergence.

\vspace{0.5cm}
\section{Higher-order Zeno sequences and Trotterization} \label{sec: High_zeno}
Quantum Zeno dynamics can be achieved in various ways \cite{Facchi_2003}. While quantum Zeno dynamics is typically obtained when a quantum system is frequently observed, the application of frequent unitary kicks and a periodic control field of high frequency can yield the same dynamics. Below, we start by developing higher-order Zeno formulas for Zeno dynamics achieved through projective measurements, followed by discussing implementations through unitary kicks and periodic control fields.

\subsection{Higher-order Zeno formulas with projective measurements}\label{projection_higher_order}

Quantum Zeno dynamics is obtained in the (Zeno) limit \cite{facchi2008quantum, zanardi2014coherent, hahn2022unification},
\begin{align}\label{Zeno_limit}
    \lim_{N \rightarrow \infty}(Pe^{-iH\Delta t}P)^N = e^{-iPHPt}P,
\end{align}
where $\Delta t = t/N$ is the time interval in which projective measurements described by the Hermitian projection $P=P^{2}$ are performed. If $P=\ket{\psi}\bra{\psi}$ is the projector onto the initial state $\ket{\psi}$, in the Zeno limit the dynamics becomes frozen as $PHP$ yields a global phase \cite{facchi2008quantum, misra1977zeno}. However, if only parts of a quantum system are frequently observed, i.e., so that the projection is of the form $P=\mathds{1}\otimes \ket{\phi}\bra{\phi}$ for some $\ket{\phi}$, in the Zeno limit the evolution takes place in the Zeno subspace \cite{Facchi_2002} that is not one-dimensional. In fact, the dynamics over the Zeno subspace can even be more complex than the dynamics over the full Hilbert space \cite{eb3d95ca9b7c45eab60553aab2734edb, Arenz_2016}.

The speed of convergence to Zeno dynamics can be determined by upper-bounding the error,
\begin{align}
\epsilon=  \norm{\left( Pe^{-iH\frac{t}{N}}P\right)^N - e^{-iPHPt}P},
\end{align}
that is obtained for finitely many projective measurements $N$ \cite{hahn2022unification}, 
\begin{align}\label{Zeno_error_1}
    \epsilon \leq \frac{t^2\norm{H}^2}{N},
\end{align}
where throughout this work $\Vert \cdot\Vert$ denotes the spectral norm. 
We note that the probability $p$ of successfully implementing the sequence $\left( Pe^{-iH\Delta t}P\right)^N$ can be lower bounded in a similar way, yielding $p \geq 1- 2\frac{\norm{H}^2 t^2}{N}$. Thus, the speed of convergence to Zeno dynamics is of the order $\mathcal O(1/N)$. This result begs the question whether Zeno dynamics can be achieved faster, i.e., with fewer measurements to achieve a given precision. For example, can the speed of convergence be improved by performing projective measurements in different time intervals, similarly to achieving faster convergence through higher-order Trotter formulas? Unfortunately, this is not the case, as terms of the form $PH^{2}P$ do not cancel to yield Zeno dynamics that is governed by $PHP$.  However, building on \cite{dhar2006preserving}, some of the authors recently showed \cite{dizaji2024hamiltoniansimulationzenosubspaces} that a second-order Zeno sequence that yields an $\mathcal O(1/N^{2})$  convergence rate can be constructed through interspersing the observed dynamics with the unitary transformation $R=P-Q$ where $P$ and $Q$ are orthogonal projections that satisfy $P+Q=\mathds{1}$.  Indeed, we showed that a sequence of the form  $\left(Pe^{-iH\Delta t/ 2} R e^{-iH\Delta t/ 2} P \right)^N$, yields a Zeno error,
\begin{align}
\epsilon=\norm{\left( Pe^{-iH\Delta t/2} R e^{-iH \Delta t/2}P\right)^N - e^{-iPHPt}P},
\end{align}
that is upper-bounded by,
\begin{align}
\epsilon\leq \frac{t^3\Vert H\Vert^3}{3N^2}. 
\end{align}
This sequence is implemented with a probability lower bounded by $p \geq 1-\frac {4\norm{H}^3 t^3}{3N^2}$.

We go on to generalize the approach described above to develop Zeno sequences that converge to Zeno dynamics at a rate $\mathcal O(1/N^{2k})$, where $k=1,2,\cdots$ denotes the order of the sequence.  
In particular, if we denote by $\mathcal{U}_{2k}(\Delta t)$ a $2k$th-order Zeno sequence that yields Zeno dynamics in the Zeno limit, 
\begin{equation}
    \lim_{N\rightarrow\infty} \left(P\mathcal{U}_{2k}(\Delta t)P\right)^N = e^{-iPHP t} P,
\end{equation}
and that satisfies,
\begin{equation}
    \norm{P\mathcal{U}_{2k}(\Delta t)P - e^{-iPHP \Delta t}P} = \mathcal{O} \left(\frac{1}{N^{2k+1}}\right).
\end{equation}
The Zeno error,
\begin{align}
\epsilon_{2k}= \norm{\left(P\mathcal{U}_{2k}(\Delta t)P\right)^{N} - e^{-iPHPt}P},
\end{align}
associated with the $2k$-order sequence can be upper-bounded by using a telescoping type argument \cite{hahn2022unification} to obtain the desired result,
\begin{align}\label{higher_order_error_1}
\epsilon_{2k}\leq N\norm{P\mathcal{U}_{2k}(\Delta t)P - e^{-iPHP \Delta t}P}=\mathcal O \left(\frac{1}{N^{2k}}\right).
\end{align}

We can construct $\mathcal{U}_{2k}(\Delta t)$ by using the theory of higher-order Trotter formulas \cite{childs2021theory,10.1063/1.529425}. This is achieved by first noting that the Hamiltonian can be written as,
\begin{align}
H=H_{Z}+H_{PQ},
\end{align}
where,
\begin{align}
H_{Z}=PHP+QHQ,
\end{align}
is the Zeno Hamiltonian that governs the Zeno dynamics, reducing to $PHP$ when the dynamics is preceded by $P$ as in \eqref{Zeno_limit}. The term,
\begin{align}
H_{PQ}=PHQ+QHP,
\end{align}
describes the coupling between the two orthogonal subspaces associated with $P$ and $Q$. 
Now, the key observation is that since,
\begin{align}
    RHR = H_Z - H_{PQ},
\end{align}
the term $H_{PQ}$ is suppressed if we rapidly alternate between the evolution generated by $H$ and $RHR$. Thus, the Trotter sequence $\left(e^{-iH\Delta t/2}Re^{-iH\Delta t/2}R\right)^{N}$ yields, in the limit $N\to\infty$, Zeno dynamics governed by $H_{Z}$. If the Trotterization is done with different time intervals, involving a longer sequence, higher-order Trotter formulas \cite{childs2021theory,10.1063/1.529425,berry2007efficient} can be used to construct higher-order Zeno formulas. Specifically, if we denote by $\mathscr{S}_{2k}$ a $2k$th-order Trotter formula for creating an evolution generated by $\mathcal{H} = H + RHR$ with Trotter step size $\Delta t/2$ (see \cite{childs2021theory} for an explicit construction), then,
\begin{equation}\label{trotter_error}
    \norm{\mathscr{S}_{2k} - e^{-iH_Z \Delta t}} = \mathcal{O}\left(\alpha( \Delta t/2)^{2k+1}\right),
\end{equation}
where,
\begin{equation} 
\label{eq:defalpha}
    \alpha= \sum_{\gamma_1,\cdots ,\gamma_{2k+1}=1}^2 \norm{[H_{\gamma_{2k+1}}, \dots,[H_{\gamma_2} , H_{\gamma_1}]]},
\end{equation}
and,
\begin{equation}
    H_{\gamma_i} = \begin{cases}
    H   \qquad &\text{if} \qquad \gamma_i = 1,\\
    RHR \qquad &\text{if} \qquad \gamma_i = 2.
    \end{cases}
\end{equation}
We consequently obtain,
\begin{align}\label{eq:higherorderscaling}
   & \norm{\left( P\mathscr{S}_{2k}P \right)^N - e^{-i PHP t}P} = \mathcal{O}\left(\alpha  \frac{t^{2k+1}}{N^{2k}}\right),
\end{align}
by a similar telescoping argument as in \eqref{higher_order_error_1}. 
Thus, a $2k$th-order Zeno sequence $\mathcal{U}_{2k}(\Delta t) = \mathscr{S}_{2k}$ can be constructed through a $2k$th-order Trotter sequence that alternates between the evolutions governed by $H$ and $RHR$ with a step size of $\Delta t/2$.  With further details found in the Appendix~\ref{Success_probability}, the success probability for implementing the sequences is of the order $p_{2k}=1-\mathcal{O} \left(\alpha^2\frac{t^{4k+2}}{N^{4k+1}}\right)$.

\subsection{Higher-order Zeno formulas with unitary kicks}\label{unitary_higher_order}

Quantum Zeno dynamics can also be realized by rapidly interspersing the dynamics with a unitary transformation $U_{Z}$ (``kick'') to obtain the Zeno sequence $\left(U_Ze^{-iH\Delta t}\right)^N$. For sufficiently large $N$, the dynamics is given by $U_Z^Ne^{-iH_Z t}$ \cite{burgarth2022one, PhysRevA.69.032314}, where $H_{Z}= \sum_{l=1}^{m} P_lHP_l$ is the Zeno Hamiltonian obtained from the eigenprojectors $P_{l}$ of $U_{Z}$ with corresponding distinct eigenvalues $\{e^{-i\phi_l}\}$. The corresponding Zeno error,
\begin{equation}
\label{eq:errorboundKicks}
\epsilon_{\text{kick}}=\left \Vert \left(U_{Z}e^{-iH\Delta t}\right)^{N}-U_{Z}^{N}e^{-iH_{Z}t} \right \Vert,
\end{equation}
that is introduced for finite $N$ can be upper bounded by \cite{burgarth2022one}, 
\begin{equation}
    \epsilon_{\text{kick}} \leq \frac{2}{N}\left(\frac{\sqrt{m}}{\eta}+1\right)t\Vert H\Vert (1+2\Vert H\Vert t ),
\end{equation}
where $\eta=\min_{k\neq l}\left |e^{-i\phi_{k}}-e^{-i\phi_{l}} \right|$.

Now, by identifying the unitary kick $U_Z$ with the reflection operator $R$ so that the Zeno Hamiltonian is given by $H_{Z}=PHP+QHQ$, the bound becomes  \cite{dizaji2024hamiltoniansimulationzenosubspaces},
\begin{align}\label{Old_bound}
   \epsilon_{\text{kick}}\leq \frac{2}{N} \left (\frac{1}{\sqrt{2}}+1\right)\norm{H}t(1+2\norm{H} t). 
\end{align}
We note that this upper bound does not depend on whether $H$ and $RHR$ commute or not. As such, in cases where $[H,RHR]=0$, the bound is not tight. The upper bound on the Zeno error $\epsilon_{\text{kick}}$ can be improved to capture the scaling with the norm of the commutator of $H=H_{Z}+H_{PQ}$ and $RHR=H_{Z}-H_{PQ}$, by observing that $(R e^{-iH\Delta t/2})^{2N} = \left(e^{-iRHR \Delta t/2} e^{-iH\Delta t/2}\right)^N$. As such, we can leverage the upper bound for the (first) Trotter formula \cite{childs2021theory,10.1063/1.529425} to obtain an improved bound for the Zeno error, 
\begin{align}
\epsilon_{\text{kick}}\leq \frac{t^{2}}{8N}\Vert[H,RHR]\Vert,
\end{align}
for Zeno dynamics achieved through unitary kicks $R$. 

Higher-order Zeno formulas can be obtained in the unitary kick case by leveraging again higher-order Trotter formulas $\mathscr{S}_{2k}$ that alternates between the evolution generated by $H$ and $RHR$ with a step size $\Delta t/2$. Analogously to the derivation that led to the scaling in \eqref{eq:higherorderscaling}, we find,
\begin{align}\label{eq:zeno_unitary_N}
 \norm{\mathscr{S}_{2k}^N - e^{-i H_Z t}} = \mathcal{O}\left(\alpha\frac{t^{2k+1}}{N^{2k}}\right),
\end{align}
where $\alpha$ is defined in \eqref{eq:defalpha}. 

\subsection{Zeno dynamics through a periodic control field of high frequency}

Achieving Zeno dynamics through unitary kicks can be viewed as a special case of a quantum system subject to a high-frequency periodic field  $\phi(t)$ that controls a projector $P$ acting on the system.  Namely, the evolution that is governed by the time-dependent Hamiltonian,
\begin{align}
\label{eq:controlledEvo}
H_{\text{tot}}(t) = H + \phi(t)P,
\end{align}
in the high-frequency limit, approximates the dynamics governed by the Zeno Hamiltonian $H_Z$~\cite{PhysRevA.71.022302}. In particular, when the control field $\phi(t)$ is chosen as a sequence of delta functions applied at discrete times (bang-bang control), the resulting dynamics reproduces the Zeno sequence corresponding to the unitary-kick formulation of the quantum Zeno effect \cite{McCaw_2005,Combescure1990}. We next discuss how Zeno dynamics can be achieved through a smooth modulation of $\phi(t)$, rather than delta-function-like pulses.

We consider the situation where a quantum system is subject to a periodic control field $\phi(t)$ with period $T$, i.e., $\phi(t+T)=\phi(t)$, whose coupling to the system is described by the Hamiltonian \eqref{eq:controlledEvo}. The operator $U(t)$ that describes the time evolution of the system up to time $t$ can be decomposed as~\cite{Lidar_Brun_2013}, 
\begin{equation}
    U(t) = U_c(t) \tilde{U}(t),
\end{equation}
where $ U_c(t) =  e^{-i \int_0^t dt'  \phi(t') P} $ describes the time evolution of the system in the absence of $H$. The unitary operator $\tilde{U}(t)=\mathcal T e^{-i\int_{0}^{t} \tilde{H}(t^{\prime})dt^{\prime}}$ is governed by the Hamiltonian $\tilde{H}(t)=U_{c}^{\dagger}(t)HU_{c}(t)$, that is given by,
\begin{align}\label{eq:tildeH}
    \Tilde{H}(t)=H_Z+e^{i\Phi(t)}PHQ+e^{-i\Phi(t)}QHP,
\end{align}
where $\Phi(t) = \int_0^t \phi(t')  dt'$ is the integrated control field. If $\Phi(T) = 2m\pi$ where $m$ is an integer, then the control propagator $U_{c}(t)$ is cyclic, i.e., $U_c(t + mT) = U_c(t)$, so that $U_{c}(mT)=\mathds{1}$. In this case, the total time evolution after a time $t=NT$ is given by,
\begin{equation}
    U(NT) = \left[\tilde{U}(T) \right]^N.
\end{equation}
We go on to express $\tilde{U}(T)=e^{-iT\Omega(T)}$ using the Magnus expansion $\Omega(T)=\sum_{n=1}^{\infty}\Omega_{n}$ which converges as long as $T<\frac{\pi}{\Vert H\Vert}$ \cite{Lidar_Brun_2013}. Since the first order of the Magnus expansion is given by $\Omega_{1}=\frac{1}{T}\int_{0}^{T}\tilde{H}(t)dt$, while higher orders scale as $\mathcal O(\Vert H\Vert^{2}T^{2})$, we see from \eqref{eq:tildeH} that if the integrated control field satisfies,
\begin{align}
\label{eq:firstorderCondition}
\int_{0}^{T}e^{\pm i \Phi(t)}dt=0,
\end{align}
the first order of the Magnus expansion becomes $\Omega_{1}=H_{Z}$ to yield Zeno dynamics. The error between the time evolution $U(t)$ at time $t$ and Zeno dynamics $e^{-iH_{Z}t}$ is of the order of,
\begin{equation}
   \Vert e^{-iH_{Z}t} -U(t)\Vert  = \mathcal{O}\left(\frac{\norm{H}^2t^2}{N}\right),
\end{equation}
showing that Zeno dynamics is obtained for control fields with sufficiently large frequency $\frac{1}{T}\propto N$ which satisfy \eqref{eq:firstorderCondition}.
This condition can, for example, be satisfied by a sinusoidal control field,
\begin{align}\label{phi-ansatz}
\phi(t)=\alpha\frac{2\pi}{T}  \sin \frac{2\pi t}{T},
\end{align}
where $\alpha$ is a root of the Bessel function of the first kind \cite{fonseca2005coherence,zhou2009quantum}.

Quantum Zeno dynamics can be achieved faster through control fields that satisfy the condition \eqref{eq:firstorderCondition} while simultaneously suppressing higher orders $\Omega_{j},~j>1$ of the Magnus expansion. More specifically, a control field that satisfies \eqref{eq:firstorderCondition} and suppresses the higher-order terms up to the $k$th-order, yields a Zeno error that is determined by the scaling $\mathcal O(\frac{\Vert H\Vert^{k+1}t^{k+1}}{N^{k}})$ of the $k$th-order $\Omega_{k}$ of the Magnus expansion.  While finding control fields that suppress all orders up to $k$ remains an open problem, since the control field \eqref{phi-ansatz} has the symmetry $\phi(T-t)=-\phi(t)$ and $\Phi(T)=0$, implying that $\tilde{H}(T-t)=\tilde{H}(t)$; thus, all even orders of the Magnus expansion vanish \cite{ng2011combining}. As such, the sinusoidal control field \eqref{phi-ansatz} yields faster convergence $\mathcal O(\frac{\Vert H\Vert^{3}t^{3}}{N^{2}})$ to Zeno dynamics (see Appendix \ref{apdx:continuous_2nd}), similar to the second-order Zeno sequences developed in the previous sections.

\tocless\section{Shorter Zeno sequences}

In general, the higher-order Zeno sequences developed in the previous sections require the application of exponentially many reflection operators $R$. Specifically, realizing a Zeno sequence of order $2k$ requires $2\cdot5^{k-1}$ reflection operators. In this section, we investigate how shorter sequences can be constructed that achieve the same Zeno error scaling while requiring fewer applications of $R$. For instance, consider the single Zeno step,
\begin{equation}\label{third_order_compact}
        Pe^{-i\alpha H \Delta t}Re^{-i\beta H \Delta t}Re^{-i\beta H \Delta t}Re^{-i\alpha H \Delta t}P,
\end{equation}
where $\alpha$ and $\beta$ satisfy,
\begin{align}
    \alpha + \beta = \frac{1}{2}, \label{eq: Third_1}\\
    8\alpha\beta^2 = \frac{1}{6}.\label{eq: Third_2}
\end{align}
These equations are approximately solved by  $\alpha = 0.675604$ and $\beta = -0.175604$,
yielding a sequence that approximates $e^{-iPHP\,\Delta t}P$ with an error scaling as $\mathcal{O}(\|H\|^{4}\Delta t^{4})$. Thus, the sequence \eqref{third_order_compact} requires $3N$ applications of $R$ when repeated $N$ times. 
However, we note that without the projective measurements, i.e., when only unitary kicks described by $R$ are used, the sequence converges to $e^{-iH_Z\Delta t}$ with an error that scales as $\mathcal{O}(\norm{H}^3\Delta t^3)$. This behavior is similar to the second-order Zeno sequence derived in \cite{dizaji2024hamiltoniansimulationzenosubspaces},
\begin{equation}
    P R e^{-iH \Delta t/2} R e^{-iH \Delta t/2} P,
\end{equation}
where without the projective measurements, the sequence yields an error scaling of $\mathcal {O} (\norm{H}^2\Delta t^2)$. We remark that the first $R$ is included in the sequence to relate the sequence with projective measurements to the sequence without projective measurements, noting that $PR = P$. Similarly, the sequence,
\begin{align}\label{fourth_order_compact}
    P&e^{-i\alpha H \Delta t}Re^{-i\beta H \Delta t}Re^{-i\gamma H \Delta t}R \nonumber \\ 
    &\times e^{-i\gamma H \Delta t}Re^{-i\beta H \Delta t}Re^{-i\alpha H \Delta t}P,
\end{align}
approximates $e^{-iPHP\Delta t}P$ with an error of $\mathcal{O}(\norm{H}^5\Delta t^5)$. It requires $5N$ applications of $R$ to achieve the same overall error scaling, $\mathcal{O}(\frac{\norm{H}^5 t^5}{N^4})$, as the fourth-order Zeno sequence developed in Section~\ref{projection_higher_order}, which requires $10N$ applications of the reflection operator. The coefficients are determined by the set of equations,
\begin{align}
    \alpha + \beta + \gamma = \frac{1}{2},\\
    8\beta(\alpha\beta + 2\alpha\gamma +\gamma^2) = \frac{1}{6},\\
    8\beta(\alpha^2 \beta + 2\alpha^2 \gamma + 2\alpha\gamma^2 + \beta \gamma^2) = \frac{1}{24}.
\end{align}

In general, higher-order Zeno sequences that yield a Zeno error scaling of $\mathcal{O}(\norm{H}^{m+2}\Delta t^{m+2})$, while requiring only $2m-1$ applications of $R$, can be constructed via a sequence of the form,

\begin{align}
W_{m}(\Delta t)=Pe^{-iH x_1 \Delta t}R \cdots Re^{-iHx_{2m} \Delta t}P.  
\end{align}
The total number of algebraic equations that must be satisfied to achieve convergence of order $m+1$ is $2^{m+1}-1$. However, not all of these equations are independent. In Appendix~\ref{apdx:third_order_compact}, we explicitly examine the third-order sequence~\eqref{third_order_compact} to analyze the set of equations that arise from the Taylor expansion of the sequence.

\vspace{0.5cm}

\section{More efficient Zeno sequence in the weak coupling limit and randomization}

In the special case where $\|H_{PQ}\|=J$ is small compared to $\|H_{Z}\|=\beta$, one can design higher-order Zeno sequences that more effectively suppress the Zeno error. In this section, we discuss two constructions and the performance of the corresponding sequences in the weak-coupling regime characterized by $\beta \gg J$.

\subsection{Higher-order Zeno sequences based on Uhrig DD}
As discussed earlier, a Zeno sequence of order $2k$ derived from Trotter formulas can suppress the Zeno error up to order $\mathcal{O}((\|H\|\Delta t)^{2k+1})$, but at the cost of requiring an exponential (in $k$) number of reflection operations.
In this section, we present a Zeno sequence inspired by the Uhrig dynamical decoupling (UDD) protocol \cite{uhrig2007keeping,yang2008universality}. In the regime $\beta \gg J$ (precisely specified below), both $2k$th-order Trotter- and $2k$th-order UDD-based Zeno sequences exhibit the same leading-order error scaling, $\mathcal{O}(J\beta^{2k}\Delta t^{2k+1})$, while our UDD-based construction requires only $2k$ reflections, providing an exponential reduction in resource cost.

\begin{figure}[t!]
    \centering
\includegraphics[width=8.6cm]{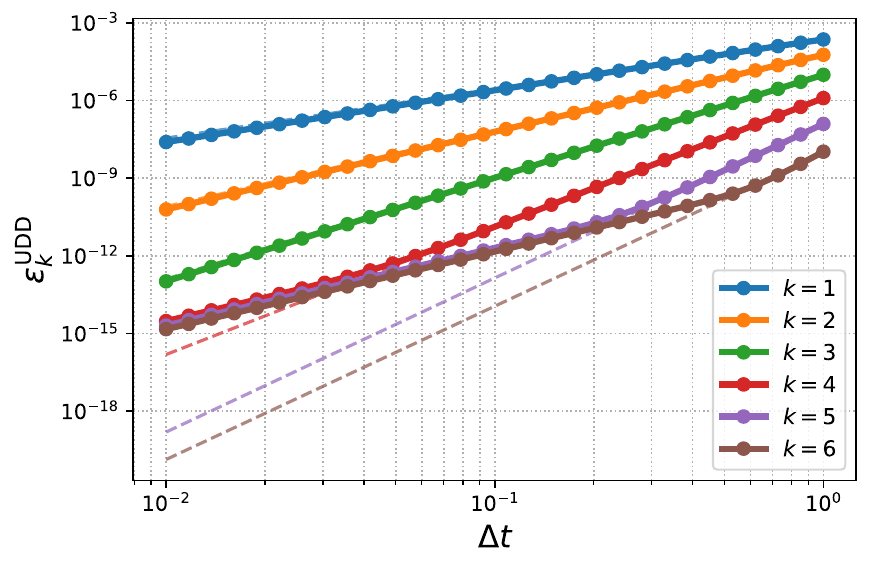}
\caption{\justifying Log–log plot of the UDD-based Zeno sequence error $\epsilon_k^{\rm{UDD}}$ given in \eqref{eq:udd_error} as a function of $\Delta t$ for different $k = 1, \cdots, 6$. The dashed lines are linear fits that have slopes $\approx k + 1$, which confirms the predicted scaling in the weak coupling regime. For smaller $\Delta t$, the error is dominated by the subleading term $\mathcal O(J^{2}\beta\Delta t^{3})$ in \eqref{eq:udd_error}, confirming a $\Delta t^3$ scaling.}
\label{fig:zeno_error_weakJ}
\end{figure}

Fix $k \ge 1$, define the time intervals $t_j = \Delta t \sin^2 \left( {j\pi}/{2(k+1)} \right)$ for $j = 0, \dots, k+1$, and consider the following sequence:
\begin{align}
\begin{split}\label{eq:udd_zeno_seq}
U_k^{\rm{UDD}}(\Delta t) 
&= e^{-i H (\Delta t - t_k)} R 
   e^{-i H (t_k - t_{k-1})} R  \\
& \qquad \cdots 
   R e^{-i H (t_2 - t_1)} R 
   e^{-i H t_1}.
\end{split}
\end{align}
Then the Zeno error,
\begin{align}
\label{eq:udd_error}
\epsilon_k^{\rm{UDD}}=\bigl\lVert U_k^{\rm{UDD}}(\Delta t)-e^{-iH_Z\Delta t}\bigr\rVert,
\end{align}
for one Zeno step is of the order of $\epsilon_k^{\rm{UDD}}= \mathcal{O}(J \beta^k \Delta t^{k+1})+ \mathcal{O}(J^2 \beta \Delta t^3)$. We refer to Appendix~\ref{apdx:udd_proof} for the details regarding the proof. The asymptotic scaling of the Zeno error for one Zeno step implies that, when $J < \beta^{k-1}\Delta t^{k-2}$, the Zeno error scales as $\mathcal O(J\beta^{k}\Delta t^{k+1})$ by using only $k$ applications of $R$.

As an example, we consider $H = \beta Z \otimes Z + J/2 (X \otimes \mathds{1}  + \mathds{1} \otimes X)$ and $P = \frac{I + Z \otimes Z}{2}$, which yields $R  = Z \otimes Z$. In this case, $\|H_Z\| = \|\beta Z \otimes Z\| = \beta$ and $\|H_{PQ}\| = \|J/2 (X \otimes \mathds{1}  + \mathds{1} \otimes X)\| = J$, where we choose $\beta = 1, J = 10^{-4}$ to satisfy the weak coupling assumption.  In Fig.~\ref{fig:zeno_error_weakJ}, we numerically investigate the Zeno error \eqref{eq:udd_error} as a function of $\Delta t$ for  $k=3,4,5,6$. For large $\Delta t$ (rightmost segments), each curve follows the predicted linear scaling with slope $\approx k+1$, demonstrating the $\mathcal{O}(J \beta^k \Delta t^{k+1})$ behavior. However, the subleading term in \eqref{eq:udd_error}, $\mathcal{O}(J^2\beta \Delta t^3)$, overtakes the first term below a crossover scale $\sim (J/\beta^{k-1})^{1/(k-2)}$, so that for fixed values of $J,\beta$, the transition into the cubic slope occurs exponentially fast.

\subsection{Randomized higher-order Zeno sequences}
In the weak-coupling regime, a simple randomization technique inspired by \cite{yi2024fasterrandomized} can improve the error scaling further. Specifically, for a quantum state $\rho$, we consider a randomized protocol where, with probability $1/2$, either $\mathscr{S}_{2k}^N$ or $R\mathscr{S}_{2k}^N R$ is applied, described by the quantum channel,
\begin{align}
\label{eq:quantumchannelrand}
\mathcal{E}_{2k}^{(N)}(\rho) =\frac12\mathscr{S}_{2k}^N\rho{\mathscr{S}_{2k}^N}^\dagger+\frac12(R\mathscr{S}_{2k}^NR)\rho(R\mathscr{S}_{2k}^NR)^\dagger.
\end{align}
Since conjugation by $R$ leaves $H_Z$ invariant while flipping the sign of $H_{PQ}$, all terms with an odd number of $H_{PQ}$’s in the error expansion cancel in $\mathcal{E}_{2k}^{(N)}$. Consequently, for $J \ll \beta$, \eqref{eq:zeno_unitary_N} yields,
\begin{align}
\left\|\mathscr{S}_{2k}^N-e^{-iH_Zt}\right\|=\mathcal{O}\left(J\beta^{2k}\frac{t^{2k+1}}{N^{2k}}\right),
\end{align}
whereas after randomization, the error is reduced to,
\begin{align}
\left\|\mathcal{E}_{2k}^{(N)}-\mathcal{U}_0\right\|_\diamond=\mathcal{O}\left(J^{2}\beta^{2k-1}\frac{t^{2k+1}}{N^{2k}}\right),
\end{align} 
where $\mathcal{U}_0(\rho)=e^{-iH_Z t}\rho e^{iH_Z t}$, and $\Vert\cdot\Vert_{\diamond}$ denotes the diamond norm. This error scaling implies that when $J$ is small compared to $\beta$, the randomization described by \eqref{eq:quantumchannelrand} yields a substantial improvement in the error scaling. In addition, it only requires two additional reflection operators. We refer to Appendix \ref{randomized_high_order} for derivations and to Appendix \ref{apdx_numeric_randomized_zeno} for the numerical investigation of the resulting advantage.

\section{Conclusion}

We developed higher-order Zeno sequences that yield faster convergence to Zeno dynamics, with an error scaling $\mathcal O\left(1/N^{2k}\right)$, where $k$ describes the order of the sequence and $N$ is the number of projective measurements performed in a fixed evolution time. This was achieved by interspersing the observed dynamics with a unitary reflection operator $R=P-Q$, where $P$ is the projector describing the projective measurement and $Q$ is its orthogonal complement. We showed that Zeno sequences constructed in this way can be related to higher-order Suzuki-Trotter formulas, which allowed us to leverage the corresponding error bounds \cite{childs2021theory, 10.1063/1.529425} to develop bounds for the Zeno error and characterize its asymptotic scaling for Zeno dynamics achieved through higher-order Zeno sequences. We discussed implementations of the developed sequences through projective measurements and unitary kicks. In the unitary kick case, we developed an improved error bound that scales with the norm of the commutator of the Hamiltonian $H$ governing the undriven evolution and $RHR$. We went on to discuss achieving Zeno dynamics faster through the application of a periodic control field of high frequency and explicitly constructed control fields that led to a second-order type error scaling.  

We further proposed alternative Zeno sequences that realize third- and fourth-order convergence toward Zeno dynamics with fewer applications of $R$. Finally, we discussed more efficient implementations by drawing relations to Uhrig dynamical decoupling and developed a randomized scheme that yields further improvements in the Zeno error scaling.

Looking ahead, several open problems merit deeper investigation. First, it remains to be determined whether there exist other classes of continuous control fields capable of generating higher-order cancellations within the Magnus expansion. Second, a systematic framework is needed for analyzing the algebraic equations that emerge from the Taylor expansions of shorter Zeno sequences, which would yield higher-order Zeno sequences that can be implemented more efficiently, i.e., with fewer applications of the reflection operator $R$.

\section*{Acknowledgements}
We acknowledge fruitful discussions with Daniel Burgarth and Nathan Wiebe. 
L.K. and M.M. acknowledge the support from the DOE (Grant No. DE-SC0024685 and DE-SC0026373). K. D. and C. A. acknowledge support from the NSF (Grant No. 2231328). 

\bibliography{Zeno.bib}
\onecolumngrid

\newpage
\appendix

\section{Success probability for implementing the higher-order Zeno sequences with projective measurements}\label{Success_probability}
We denote by $\tilde{p}_{2k}$ the probability of successfully implementing a single Zeno step of a Zeno sequence of order $2k$, satisfying,\begin{align}\mathscr{S}_{2k} = e^{-iH_Z\Delta t} + \mathcal{O}(\alpha\Delta t^{2k+1}).
\end{align}
Assuming that the initial state lies in the Zeno subspace, i.e., $P\ket{\psi} = \ket{\psi}$ and $Q\ket{\psi} = 0$, the probability $\tilde{p}_{2k}$ is given by,
\begin{equation}
    \tilde{p}_{2k} = \bra{\psi} \mathscr{S}_{2k}^{\dagger} P \mathscr{S}_{2k} \ket{\psi} = \bra{\psi} P\mathscr{S}_{2k}^{\dagger} P \mathscr{S}_{2k} P\ket{\psi},
\end{equation}
where substituting $P = \mathds{1} - Q$ yields, 
\begin{align}
    \tilde{p}_{2k} &= 1 - \bra{\psi} (P\mathscr{S}_{2k}^{\dagger} Q )(Q \mathscr{S}_{2k} P)\ket{\psi}\\
    &\geq 1- \norm{(P\mathscr{S}_{2k}^{\dagger} Q ) (Q \mathscr{S}_{2k} P)}\\
    &\geq 1- \lVert P\mathscr{S}_{2k}^{\dagger} Q\rVert  \norm{Q \mathscr{S}_{2k} P}.
\end{align}
Since $\mathscr{S}_{2k} = e^{-iH_Z\Delta t} + \mathcal{O}(\alpha\Delta t^{2k+1})$, we have $\lVert P\mathscr{S}_{2k}^{\dagger} Q\rVert = \norm{Q \mathscr{S}_{2k} P} = \mathcal{O}(\alpha \Delta t^{2k+1})$, which implies,
\begin{equation}
    \tilde{p}_{2k} = 1 -\mathcal{O}(\alpha^2\Delta t^{4k+2}).
\end{equation}
Therefore, the total success probability $p_{2k} = \tilde{p}_{2k}^N$ for implementing a sequence of $N$ Zeno steps scales as,
\begin{equation}
    p_{2k} = 1- \mathcal{O}\left(\alpha^2\frac{t^{4k+2}}{N^{4k+1}}\right),
\end{equation}
where $t=N\Delta t$.

\section{Time-symmetric continuous control}\label{apdx:continuous_2nd}
Assume the control field $\phi(t)$ satisfies $\phi(T - t)=-\phi(t)$. Then,
\begin{align}
\Phi(T)=\int_{0}^{T}\phi(s)ds =\int_{0}^{T/2}\big[\phi(s)+\phi(T-s)\big]ds=0,
\label{eq:B1}
\end{align}
which yields, 
\begin{align}
\Phi(T-t)=\int_{0}^{T-t}\phi(s)ds =\int_{t}^{T}\phi(T-u)du =-\int_{t}^{T}\phi(u)du =-(\Phi(T)-\Phi(t))=\Phi(t),
\end{align}
to arrive at,
\begin{align}\label{eq:B2}
\Phi(T - t)=\Phi(t).
\end{align}
Using \eqref{eq:tildeH} in the main text and \eqref{eq:B2} we find, 
\begin{align}
\tilde H(T - t)=H_Z + e^{+i\Phi(T-t)}PHQ + e^{-i\Phi(T-t)}QHP=\tilde H(t),
\label{eq:B3}
\end{align}
which shows that $\tilde H(t)$ is time-symmetric on $[0,T]$. For such time-symmetric generators, it is a standard result that all even orders in the Magnus expansion over $[0,T]$ vanish (e.g., see Ref.\cite{ng2011combining}), i.e.
\begin{align}\label{eq:B4}
\Omega_{2k}=0\quad (k=1,2,\ldots).
\end{align}
The sinusoidal control of \eqref{phi-ansatz},
\begin{align}
\phi(t)=\alpha\frac{2\pi}{T}\sin\Big(\frac{2\pi t}{T}\Big), 
\label{eq:B5}
\end{align}
satisfies the symmetry relation $\phi(T-t) = -\phi(t)$. The first-order cancellation condition
$\int_0^{T} e^{\pm i\Phi(t)}dt=0$ gives,
\begin{align}
\int_{0}^{T}e^{\pm i\Phi(t)}dt =\frac{T}{2\pi}\int_{0}^{2\pi} e^{\pm i\alpha(1-\cos\theta)} d\theta =Te^{\pm i\alpha}J_0(\alpha)=0,
\label{eq:B6}
\end{align} where $J_0$ is the Bessel function of the first kind. Hence, choosing $\alpha$ to be a zero of $J_0$ (e.g., $\alpha \approx 2.4$) yields $\Omega_1=H_Z$. Together with \eqref{eq:B4}, the leading error is $T\Omega_3=\mathcal O(\|H\|^3T^3)$, and over $t=NT$ this gives the scaling $\mathcal O(\|H\|^{3}t^{3}/N^{2})$, as given in the main text.

\section{Third-order compact sequence}\label{apdx:third_order_compact}
To analyze the equations that arise from the Taylor expansion of the sequence~\eqref{third_order_compact}, we expand each exponential up to third order in $\Delta t$:
\begin{align}
    P&e^{-i\alpha H \Delta t}Re^{-i\beta H \Delta t}Re^{-i\beta H \Delta t}Re^{-i\alpha H \Delta t}P = \\
    P&\Bigl(\mathds{1}- i\alpha H\Delta t - \frac{\alpha^2\Delta t^2}{2} H^2 +i\frac{\alpha^3\Delta t^3}{6}H^3 + \mathcal{O}(\norm{H}^4\Delta t^4)\Bigl)(2P-\mathds{1})\\
    &\Bigl(\mathds{1}- i\beta H\Delta t - \frac{\beta^2\Delta t^2}{2} H^2 +i\frac{\beta^3\Delta t^3}{6}H^3 + \mathcal{O}(\norm{H}^4\Delta t^4)\Bigl)(2P-\mathds{1})\\
    &\Bigl(\mathds{1}- i\beta H\Delta t - \frac{\beta^2\Delta t^2}{2} H^2 +i\frac{\beta^3\Delta t^3}{6}H^3 + \mathcal{O}(\norm{H}^4\Delta t^4)\Bigl)(2P-\mathds{1})\\
    &\Bigl(\mathds{1}- i\alpha H\Delta t - \frac{\alpha^2\Delta t^2}{2} H^2 +i\frac{\alpha^3\Delta t^3}{6}H^3 + \mathcal{O}(\norm{H}^4\Delta t^4)\Bigl)P.
\end{align}
We then separate the different powers of $\Delta t$:
\begin{align}
    \text{Zeroth order:}&\quad P\\
    \text{First order:}&\quad -i\Delta t(2\alpha+2\beta)PHP\\
    \text{Second order:}&\quad -2\Delta t^2(\alpha^2 + 2\alpha\beta+\beta^2)(PHP)^2\\
    \text{Third order:}&\quad +i\Delta t^3\Bigl((\alpha^3+3\alpha^2\beta-3\alpha \beta^2+\beta^3)(PHPH^2P+PH^2PHP) -  \\
    &\hspace{40pt}\frac{2}{3}(\alpha^3 +3\alpha^2\beta -3\alpha\beta^2 + \beta^3) (PH^3P) +8\alpha \beta^2 (PHP)^3 \Bigl). \nonumber
\end{align}
It is readily observed that the second-order condition is automatically satisfied given the fulfillment of the first-order constraint. Moreover, all undesired terms at the third order possess identical coefficients. In addition, if the relation $ 8\alpha\beta^2 = \frac{1}{6} $ holds, these third-order terms vanish.

\section{Proof of UDD error bound} \label{apdx:udd_proof}

Fix an integer $k \ge 1$, and define the switching times $t_j = \Delta t \sin^2 \left( \frac{j\pi}{2(k+1)} \right)$ for $j = 0,\dots,k+1$. Let $F_k(t) = (-1)^j$ for $t$ in $(t_j, t_{j+1})$, and consider the piecewise-constant Hamiltonian,
\begin{align}\label{eq:lemma_ht}
H(t) = H_Z + F_k(t)  H_{PQ}, \qquad 0 \le t \le \Delta t,
\end{align}
that generates the time evolution operator,
\begin{align}
U_k^{\rm{UDD}}(\Delta t) = \mathcal T\exp\left( -i \int_0^{\Delta t} H(t)  dt \right).
\end{align}
Then the following estimate holds,
\begin{align}
\bigl\| U_k^{\rm{UDD}}(\Delta t) - e^{-i H_Z \Delta t} \bigr\| = \mathcal{O}\bigl(J \beta^{k} \Delta t^{k+1} \bigr) + \mathcal{O}\bigl( J^2 \beta \Delta t^{3} \bigr),
\end{align}
where we recall that $\beta = \|H_Z\|$ and $J = \|H_{PQ}\|$.

To prove the above asymptotic scaling, we first move to the interaction picture with respect to $H_Z$, and write $H_{PQ}^{(I)}(t) = e^{i H_Z t} H_{PQ} e^{-i H_Z t}$, which satisfies $\| H_{PQ}^{(I)}(t) \| = J$. Then,
\begin{align}\label{udd_int_V}
U_k^{\rm{UDD}}(\Delta t) = e^{-i H_Z \Delta t} V(\Delta t), \qquad V(\Delta t) = \mathcal T \exp\left(-i \int_0^{\Delta t} F_k(t) H_{PQ}^{(I)}(t) dt \right).
\end{align}
Expanding $V$ through the Dyson series $V(\Delta t) = \sum_{n \ge 0} \Delta_n$ gives,
\begin{align}
\Delta_n = (-i)^n\int_{0 < t_1 < \cdots < t_n < \Delta t}
\left(\prod_{j=1}^{n} F_k(t_j) \right) H_{PQ}^{(I)}(t_n) \cdots H_{PQ}^{(I)}(t_1) dt_1 \cdots dt_n .
\end{align}
If we substitute the series $H_{PQ}^{(I)}(t) = \sum_{p \ge 0} C_p t^p$ with $C_p = \frac{i^{p}}{p!} \operatorname{ad}_{H_Z}^{p}(H_{PQ})$ (where $\|C_{p}\| \leq 2^p \beta^p J/p!$), and make a change of variables $t_j = \Delta t \tau_j$, we find,
\begin{align}
\Delta_n = (-i)^n \sum_{\vec p \ge 0} C_{p_n} \cdots C_{p_1} (\Delta t)^{n + \sum_j p_j} F_{\vec p},
\end{align}
where,
\begin{align}
F_{\vec p} = \int_{0 < \tau_1 < \cdots < \tau_n < 1}
\left( \prod_{j=1}^{n} F_k(\Delta t  \tau_j) \tau_j^{p_j} \right) d\tau_1 \cdots d\tau_n .
\end{align}
\medskip\noindent
We first analyze the contributions from odd Dyson orders. It was shown in \cite{yang2008universality} that $F_{\vec p} = 0$ for every odd $n \le k$ whenever $n + \sum_{j=1}^{n} p_j \le k$, so all such odd orders disappear. The first contributing odd term occurs at $n = 1$ with $p_1 = k$:
\begin{align}
\Delta_1 = (-i) C_k (\Delta t)^{k+1} \int_0^{1} F_k(\Delta t \tau)\tau^{k} d\tau.
\end{align}
The integral is $\mathcal{O}(1)$ because $|F_k| \le 1$, which yields,
\begin{align}
\| \Delta_1 \| = \mathcal{O}\bigl( J \beta^{k} (\Delta t)^{k+1} \bigr).
\end{align}

\medskip\noindent
We next analyze the even Dyson orders. The lowest surviving even-order term is $n=2$, $p_1+p_2=0$, which gives $p_1 = p_2 = 0$. However, we note that $F_{p_1=0,p_2=0} = 0$, as,
\begin{align}
F_{0,0} = \frac{1}{2} \int_0^{1} \int_0^{1} F_k(\Delta t \tau_2) F_k(\Delta t \tau_1)  d\tau_1 d\tau_2 = \frac{1}{2} \left(\int_0^{1} F_k(\Delta t  \tau) d\tau \right)^2 = 0,
\end{align}
because the average of $F_k$ on $[0, \Delta t]$ is zero for all $k \ge 1$. The lowest non-zero even contribution is therefore $n=2$ with $p_1 + p_2 = 1$:
\begin{align}
\Delta_2 = (-i)^2 (\Delta t)^{3} \sum_{p_1 + p_2 = 1} C_{p_2} C_{p_1} F_{p_1, p_2},
\end{align}
and the bounds $\| C_0 \| \leq J$, $\| C_1 \| \leq 2\beta J$, and $|F_{0,1}|, |F_{1,0}| = \mathcal{O}(1)$ give,
\begin{align}
\| \Delta_2 \| = \mathcal{O}\left( J^{2} \beta (\Delta t)^{3} \right).
\end{align}
Combining the leading odd and even contributions, and using $\| V(\Delta t) - I \| \le \sum_{n \ge 1} \|\Delta_n\|$, gives,
\begin{align}
\left\| U_k^{\rm{UDD}}(\Delta t) - e^{-i H_Z \Delta t} \right\| = \left\| V(\Delta t) - I \right\| = \mathcal{O}\left(J \beta^{k} \Delta t^{k+1} \right) + \mathcal{O}\left(J^2 \beta \Delta t^3 \right).
\end{align}
Finally, the piecewise-constant Hamiltonian in \eqref{eq:lemma_ht} generates the sequence described in \eqref{eq:udd_zeno_seq}, which directly yields \eqref{eq:udd_error}. 

We note that the ``pure UDD'' error scaling exhibits $\mathcal{O}((\| H \| \Delta t)^{k+1})$ (where $H = H_Z + H_{PQ}$), i.e., no $\Delta t^3$ term is present \cite{uhrig2007keeping,yang2008universality}. This holds when (i) the error is evaluated on the subsystem (of the protected subspace), and (ii) the subsystem parts of the preserved and orthogonal subspaces commute. Specifically, let $H_Z=A \otimes B$ and $H_{PQ} = A' \otimes B'$, where $A$ and $A'$ are Pauli operators with $[A,A'] = 0$. In this case, the interaction-picture unitary in \eqref{udd_int_V} simplifies to,
\begin{align}
    V(\Delta t)=\mathds{1}+\sum_{a = \pm 1} P_a\otimes B^{(a)} +\mathcal{O}((\| H \|\Delta t)^{k+1}),
\end{align} 
where $P_{\pm} = (\mathds{1}\pm A)/2$ are the projectors onto the subsystem $A$, and $B^{(a)}$ represents the operator acting on the second subsystem, arising from the even-order Dyson-series contributions. If the initial state is supported in a single eigenspace $P_a$, tracing out the other subsystem leaves the system unchanged up to $\mathcal{O}((\|H\|\Delta t)^{k+1})$. In the standard UDD setup (effectively $P_a=\mathds{1}$) \cite{uhrig2007keeping,yang2008universality}, this applies to arbitrary initial system states. In this regime, the sequence thus stabilizes the initial state without inducing a non-trivial evolution.

\section{Randomized higher-order Zeno sequence}\label{randomized_high_order}
\subsection{Mixing Lemma}
The mixing lemma \cite{hastings2017turning,campbell2017shorter,chen2021concentration} provides a bound on how accurately a unitary operation can be approximated by a quantum channel that describes the random application of unitary transformations. Let $\{U_j\}$ be a set of unitary matrices that are sampled according to the probability distribution $\{p_j\}$, and define the quantum channels,
\begin{align}
\mathcal{E}(\rho) = \sum_{j=0}^{N-1}p_j U_j \rho U_j^\dagger,\qquad \mathcal{V}(\rho) = V \rho V^\dagger,
\end{align}
where $V$ is unitary. Then the (sharpened) mixing lemma \cite{chen2021concentration} states that, 
\begin{align}\label{eq:mixing_sharpened}
\|\mathcal{E} - \mathcal{V}\|_{\diamond} \le 2\left\|\sum_{j=0}^{N-1}p_j U_j - V\right\|.
\end{align}

\subsection{Randomized higher-order Zeno sequence} \label{lem:rand}
We adopt the notation of the main text for $P,Q,R,H_Z,H_{PQ},J=\|H_{PQ}\|,\beta=\|H_Z\|$, and $\mathscr{S}_{2k}^{N}$. We will show that the quantum channel,
\begin{align}
\mathcal{E}_{2k}^{(N)}(\rho)=\frac12 \mathscr{S}_{2k}^{ N}\rho (\mathscr{S}_{2k}^{ N})^\dagger+\frac12 (R\mathscr{S}_{2k}^{ N}R)\rho(R\mathscr{S}_{2k}^{ N}R)^\dagger,
\end{align}
obeys,
\begin{equation}\label{eq:rand_goal}
\bigl\|\mathcal{E}_{2k}^{(N)}-\mathcal{U}_0\bigr\|_\diamond
=\mathcal{O} \left(J^{2}\beta^{ 2k-1} \frac{t^{ 2k+1}}{N^{ 2k}}\right),
\qquad \mathcal{U}_0(\rho)=e^{-iH_Z t}\rho e^{iH_Z t},
\end{equation}
i.e., randomization cancels all terms odd in $H_{PQ}$ and makes the leading dependence quadratic rather than linear in $J$ while preserving the $1/N^{2k}$ rate.

Since $RH_ZR=H_Z$ and $R=R^\dagger$ is unitary, it follows that,
\begin{align}\label{eq:norm_eq}
\bigl\|R \mathscr{S}_{2k}^{ N}R-e^{-iH_Z t}\bigr\| =\bigl\|\mathscr{S}_{2k}^{ N}-R e^{-iH_Z t} R\bigr\| =\bigl\|\mathscr{S}_{2k}^{ N}-e^{-iH_Z t}\bigr\|.
\end{align}
Write $H=H_Z+H_{PQ}$ and $RHR=H_Z-H_{PQ}$ and partition $[0,t]$ into $N$ equal intervals ($\Delta t=t/N$). Then $\mathscr{S}_{2k}^{ N}$ can be represented as a time-ordered exponential driven by piecewise-constant controls $G_N(s), F_N(s)\in\{\pm1\}$ (switching only at the $2k$th-order Trotter slices inside each subinterval), with,
\begin{align}
\int_0^{T} G_N(s) ds=t,\qquad
\mathscr{S}_{2k}^{N}=\mathcal{T}\exp \left\{-i\int_0^{T} \bigl[G_N(s)H_Z+F_N(s)H_{PQ}\bigr] ds\right\}.
\end{align}
where $T$ is the total physical duration of the sequence, and negative-time segments in the higher-order Trotter formula are implemented as forward evolution under $\pm H$ or $\pm RHR$, encoded in the signs of $G_N(s)$ and $F_N(s)$. 

Moving to the interaction picture with respect to $G_N(s)H_Z$ gives,
\begin{align}
\mathscr{S}_{2k}^{N}
& = e^{-iH_Z t} \mathcal{T}\exp \left\{-i\int_{0}^{T} F_N(s) U_Z^\dagger(s)H_{PQ}U_Z(s) ds\right\} \\
& = e^{-iH_Z t}\left[I - i \int_{0}^{T} F_N(s) H_{PQ}^{(I)}(s) ds + \mathcal{O} \Bigl(J^{2}\beta^{2k-1} \frac{t^{2k+1}}{N^{2k}}\Bigr)\right],
\end{align}
where $U_Z(s)=\mathcal{T}e^{-i\int_{0}^{s}G_N(u)H_Z du}$ and $H_{PQ}^{(I)}(s)=U_Z^\dagger(s)H_{PQ}U_Z(s)$. Here, we have isolated the term linear in $H_{PQ}$; since the underlying step is a symmetric $2k$th-order product formula, the remaining error is a sum of $(2k{+}1)$-fold nested commutators with at least two insertions of $H_{PQ}$, and hence the remainder scales as $\mathcal{O}\bigl(J^{2}\beta^{2k-1} t^{2k+1}/N^{2k}\bigr)$.

Because $[U_Z(s),R]=0$ and $RH_{PQ}R=-H_{PQ}$, conjugation by $R$ flips the sign of the linear (odd-in-$H_{PQ}$) term:
\begin{align}
R \mathscr{S}_{2k}^{N}R
= e^{-iH_Z t}\left[I + i \int_{0}^{T} F_N(s) H_{PQ}^{(I)}(s) ds + \mathcal{O} \Bigl(J^{2}\beta^{2k-1} \frac{t^{2k+1}}{N^{2k}}\Bigr)\right],
\end{align} where the remainder is the same as for $\mathscr{S}_{2k}^{N}$ as we see in \eqref{eq:norm_eq}.

Thus, averaging cancels the entire $\mathcal{O}(J\beta^{2k} t^{2k+1}/N^{2k})$ contribution in total time $t$:
\begin{align}
\left\|\frac12 \mathscr{S}_{2k}^{N}+\frac12 R\mathscr{S}_{2k}^{N}R-e^{-iH_Z t}\right\| = \mathcal{O} \left(J^{2}\beta^{2k-1} \frac{t^{2k+1}}{N^{2k}}\right).
\end{align}
Finally, for the channel $\mathcal{E}_{2k}^{(N)}(\rho)$ defined above, the (sharpened) mixing lemma in \eqref{eq:mixing_sharpened} yields,
\begin{align}
\bigl\|\mathcal{E}_{2k}^{(N)}-\mathcal{U}_0\bigr\|_\diamond
=\mathcal{O} \Bigl(J^{2}\beta^{ 2k-1} \frac{t^{2k+1}}{N^{2k}}\Bigr),
\qquad \mathcal{U}_0(\rho)=e^{-iH_Z t}\rho e^{iH_Z t}.
\end{align}

\subsection{Additional numerical simulations}\label{apdx_numeric_randomized_zeno}
\begin{figure}[htbp]
  \centering
  \begin{subfigure}[b]{0.45\textwidth}
    \centering
    \includegraphics[width=\linewidth]{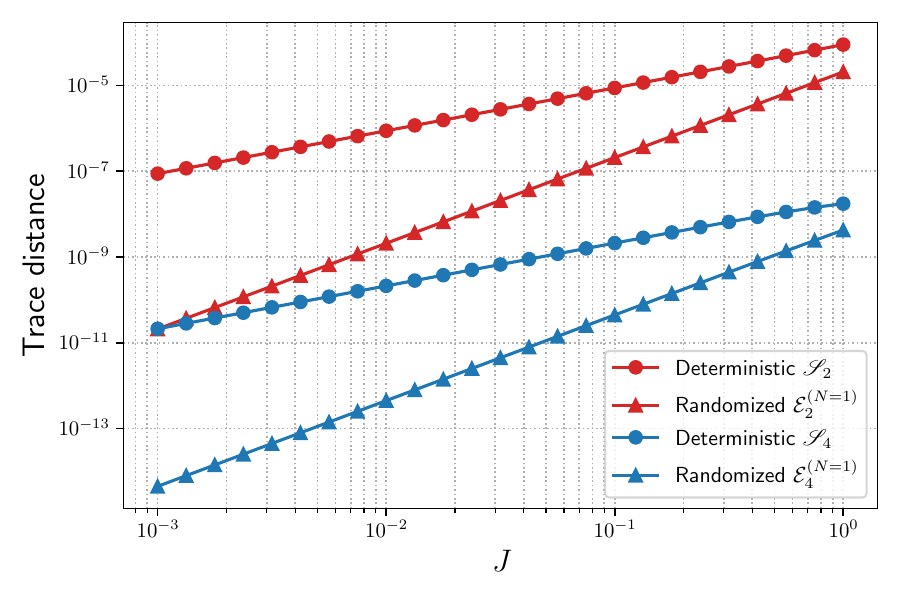}
    \label{fig:rand_J}
  \end{subfigure}
  \hfill
  \begin{subfigure}[b]{0.45\textwidth}
    \centering
    \includegraphics[width=\linewidth]{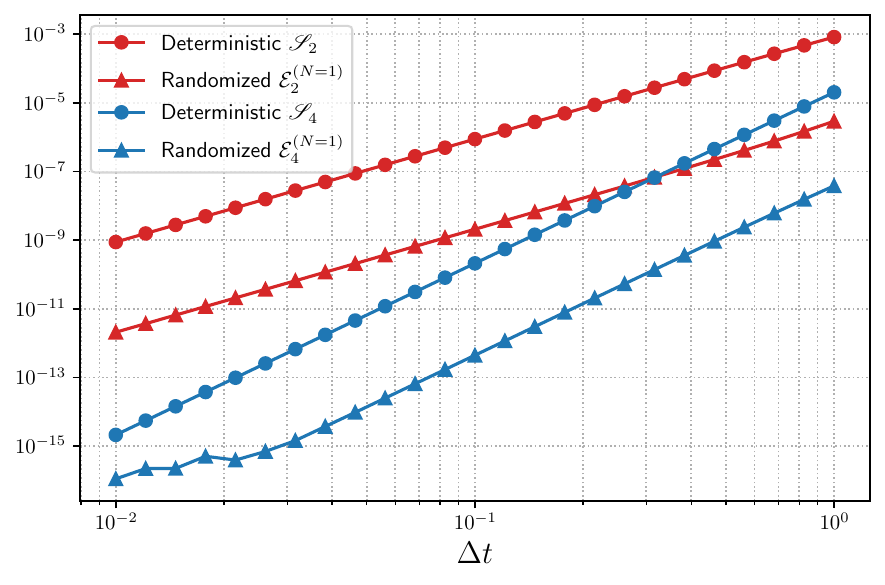}
    \label{fig:rand_dt}
  \end{subfigure}
  \caption{\justifying Trace distance error between the ideal state $e^{-iH_Z\Delta t}\rho_0 e^{iH_Z\Delta t}$ and the states produced by the deterministic $2k$th-order Zeno sequence $\mathscr{S}_{2k}$ (circles) and its randomized counterpart $\mathcal{E}_{2k}^{(N=1)}$ (triangles) for $k = 1, 2$. Deterministic errors scale as $\mathcal{O}(J\beta^{2k}\Delta t^{2k+1})$, whereas randomized errors scale as $\mathcal{O}(J^2\beta^{2k-1}\Delta t^{2k+1})$. (Left) For fixed $\Delta t = 0.1$ and varying $J$, the observed slopes confirm these scalings. (Right) For fixed $J = 0.01$, both protocols show identical $\Delta t$-scaling behavior.}
  \label{fig:rand}
\end{figure}

We consider the two-qubit Hamiltonian $H = \beta Z \otimes Z + \frac{J}{2}(X \otimes \mathds{1} + \mathds{1} \otimes X)$ together with the projector $P = \frac{1}{2}(\mathds{1} + Z \otimes Z)$, which defines the reflection operator $R = 2P - \mathds{1} = Z \otimes Z$. Throughout this example, we set $\beta = 1$, so that $\|H_Z\| = \|\beta Z \otimes Z\| = 1$ and $\|H_{PQ}\| = \|\frac{J}{2}(X \otimes \mathds{1} + \mathds{1} \otimes X)\| = J$. We compare the deterministic $2k$th-order Zeno protocols $\mathscr{S}_{2k}$ with their randomized counterparts $\mathcal{E}_{2k}^{(N=1)}$ for $k = 1, 2$. The error is quantified by the trace distance, $\frac{1}{2}\|\rho_{\mathrm{ideal}} - \rho_{\mathrm{det}}\|_1$ and $\frac{1}{2}\|\rho_{\mathrm{ideal}} - \rho_{\mathrm{rand}}\|_1$, where $\rho_{\mathrm{ideal}} = e^{-i H_Z \Delta t} \rho_0e^{i H_Z \Delta t}$, $\rho_{\mathrm{det}}= \mathscr{S}_{2k} \rho_0 \mathscr{S}_{2k}^\dagger$, and $\rho_{\mathrm{rand}}= \mathcal{E}_{2k}^{(N=1)}\left(\rho_0\right)$ with $\rho_0 = (|0\rangle \langle 0 |)^{\otimes 2}$.

In Fig.\ref{fig:rand}, the left panel shows that for a fixed time step $\Delta t = 0.1$ and varying coupling strength $J$, the error of the deterministic sequence $\frac{1}{2}\|\rho_{\mathrm{ideal}} - \rho_{\mathrm{det}}\|_1$, scales linearly with $J$, whereas the error of the randomized sequence, $\frac{1}{2}\|\rho_{\mathrm{ideal}} - \rho_{\mathrm{rand}}\|_1$, exhibits quadratic scaling. The right panel illustrates that for fixed $J = 0.01$ and varying $\Delta t$, both protocols display the same $\Delta t$-scaling behavior. Both plots are consistent with \eqref{eq:rand_goal} for $N = 1$.

\end{document}